\documentclass{iaus}
\usepackage{graphicx}

\title[JD 11.~~Monitoring of the megamaser in MG~J0414+0534] 
{Long term Arecibo monitoring of the water megamaser in MG~J0414+0534}

\author[Paola Castangia et al.]   
{Paola Castangia$^1$
 M. C. Violette Impellizzeri$^2$
 \thanks{Speaker}
 John P. McKean$^3$
 Christian Henkel$^4$
 Andreas Brunthaler$^4$
 Alan L. Roy$^4$
\and Olaf Wucknitz$^5$}

\affiliation{$^1$INAF-Osservatorio Astronomico di Cagliari, \\ Loc. Poggio dei Pini,
Strada 54, 09012 Capoterra (CA), Italy \\ email: {\tt pcastang@oa-cagliari.inaf.it} \\[\affilskip]
$^2$ALMA, \\ Chile \\email: {\tt vimpelli@alma.cl}\\[\affilskip]
$^3$ASTRON, \\ Oude Hoogeveensedijk 4, 7991 PD Dwingeloo, the Netherlands\\[\affilskip]
$^4$Max-Planck-Institut f\"ur Radioastronomie, \\ Auf dem H\"ugel 69, D-53121 Bonn, Germany\\[\affilskip]
$^5$Argelander-Institut f\"ur Astronomie, \\ Auf dem H\"ugel 71, D-53121 Bonn, Germany}

\pubyear{2012}
\volume{287}  
\pagerange{119--126}
\jname{Cosmic masers: from OH to H$_0$}
\editors{A.C. Editor, B.D. Editor \& C.E. Editor, eds.}
\begin{document}

\maketitle

\begin{abstract}
We monitored the 22\,GHz maser line in the lensed quasar MG~J0414+0534 at z=2.64 with the 300-m Arecibo
telescope for almost two years to detect possible additional maser components and to measure a potential
velocity drift of the lines. The main maser line profile is complex and can be resolved into a number
of broad features with line widths of 30-160\,km\,s$^{-1}$. A new maser component was tentatively detected in
October 2008 at a velocity of +470\,km\,s$^{-1}$. After correcting for the estimated lens magnification, 
we find that the H$_2$O isotropic luminosity of the maser in MG~J0414+0534 is $\sim$26,000 solar luminosities, 
making this source the most luminous ever discovered. Both the main line peak and continuum flux densities are
surprisingly stable throughout the period of the observations. An upper limit on the velocity drift of the
main peak of the line has been estimated from our observations and is of the order of 2\,km\,s$^{-1}$ per year. We
discuss the results of the monitoring in terms of the possible nature of the maser emission,
associated with an accretion disk or a radio jet. This is the first time that such a study
is performed in a water maser source at high redshift, potentially allowing us to study the parsec-scale
environment around a powerful radio source at cosmological distances.
\end{abstract}

\firstsection 
\section{Introduction}
To date, most surveys searching for extragalactic water masers have targeted spiral galaxies and radio-quiet AGN in the local Universe. Likely, this is the reason why, with the exception of a type 2 quasar at $z=0.66$ (\cite[Barvainis \& Antonucci 2005]{barvainis05}), the majority of the known extragalactic water masers have been found in Seyfert 2 or LINER galaxies at low redshift ($z<0.06$). Observations of objects at higher redshifts are limited in part by the range of frequencies available but mainly by the sensitivity of current radio telescopes. To overcome this limitation, we have been carrying out a survey of water masers in known gravitationally lensed quasars with the Effelsberg and Arecibo telescopes (\cite[McKean et al. 2011; McKean et al. in prep]{mckean2011}). Observing gravitationally lensed quasars allows us to use the magnification provided by the lensing galaxy to increase the measured flux density of the background AGN, strongly reducing the integration time necessary to detect the signal originating from distant objects ($z>1$). This potentially allows us to discover water masers at cosmological distances and to study the parsec-scale environment of AGN in the early Universe.
Our first confirmed high-redshift water maser was found toward the lensed quasar MG~J0414+0534 at $z=2.64$ (\cite[Impellizzeri et al. 2008]{impellizzeri08}). The line was originally detected with the Effelsberg radiotelscope and subsequently confirmed with the EVLA that found the emission to be coincident with the lensed images of the quasar (A1 and A2). The apparent isotropic luminosity of the line ($\sim$10,000\,L$_{\odot}$, after correcting for the estimated lens magnification) makes the water maser in MG~J0414+0534 not only the most distant but also one of the most luminous water masers ever detected and suggests that the emission is associated with the AGN. Based on the large line width of the Effelsberg and EVLA spectra, our initial hypothesis on the origin of the maser was that the emission is associated with the jet(s) of the radio loud quasar. In order to reveal possible variations in the maser flux density, typical of the masers produced by the interactions between a molecular cloud and a radio jet, and determine if a correlation exists between the maser and the continuum emission, we monitored the line and the radio continuum in MG~J0414+0534 with the 300-m Arecibo telescope during a time interval of 15 months. Here we report the results of this monitoring campaign and discuss them within the framework of the two main scenarios for the origin of the maser, i.\,e. jet-maser and disk-maser emission (for more details on this work, see \cite[Castangia et al. 2011]{castangia2011}). We adopt a cosmology with $\Omega_{\rm M} =0.3$, $\Omega_{\rm \Lambda} =0.7$ and $H_0 = 70$\,km\,s$^{-1}$\,Mpc$^{-1}$. In the following, the quoted line velocities are defined w.r.t. the optical redshift of MG~J0414+0534 ($z$\,=\,2.639; \cite[Lawrence et al. 1995]{lawrence95}), using the optical velocity definition in the heliocentric frame. 

\section{Results}
We have monitored the redshifted (rest frequency: 22\,GHz) radio continuum and maser emission in MG~J0414+0534 for $\sim$15 months at $\sim$6 week intervals and found that both are surprisingly stable. Absolute deviations of the continuum flux from the mean are on average comparable with the flux calibration uncertainty (7\%). The 6\,GHz (observed frequency) continuum flux density of MG~J0414+0534 thus remained nearly constant for the duration of the entire monitoring period, with an average flux density of 0.71$\pm$0.02\,Jy, the error being the standard deviation of the mean. The line peak flux density is also surprisingly stable throughout the period of the observations. Small fluctuations are not exceeding the limits of uncertainty (between 10\% and 50\%). From the analysis of the 11 epochs of the monitoring, we can place an upper limit on the velocity drift of the line peak of 2\,km\,s$^{-1}$\,yr$^{-1}$. 

\begin{figure}[b]
\begin{center}
 \includegraphics[angle=90,width=3.4in]{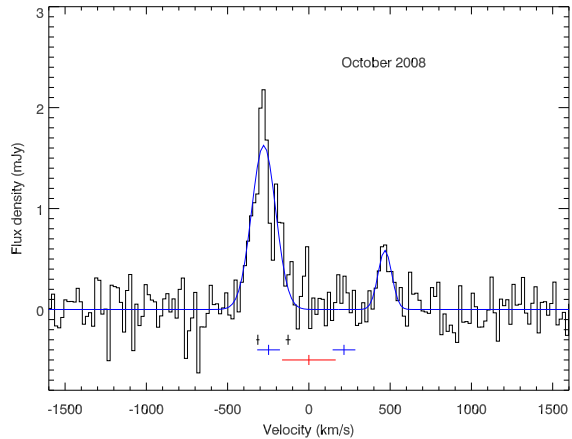} 
 \caption{Water maser spectrum observed towards MG~J0414+0534 in October 2008 (black histogram). The fitted Gaussian profiles are overlaid (blue line). The channel spacing is 19.2\,km\,s$^{-1}$. The root-mean-square (r.m.s.) noise level of the spectrum is 0.2\,mJy per channel. The velocity scale is relative to redshift 2.639 using the optical velocity definition in the heliocentric frame. The red cross marks the systemic velocity and the associated uncertainty. The blue and the black crosses indicate the peaks of the CO emission (\cite[Barvainis et al. 1998]{barvainis98}) and the H{\sc i} absorption components (\cite[Moore et al. 1999]{moore99}), respectively, with their errors.}
   \label{fig:fig_oct}
\end{center}
\end{figure}

\begin{figure}[b]
\begin{center}
 \includegraphics[angle=-90,width=3.4in]{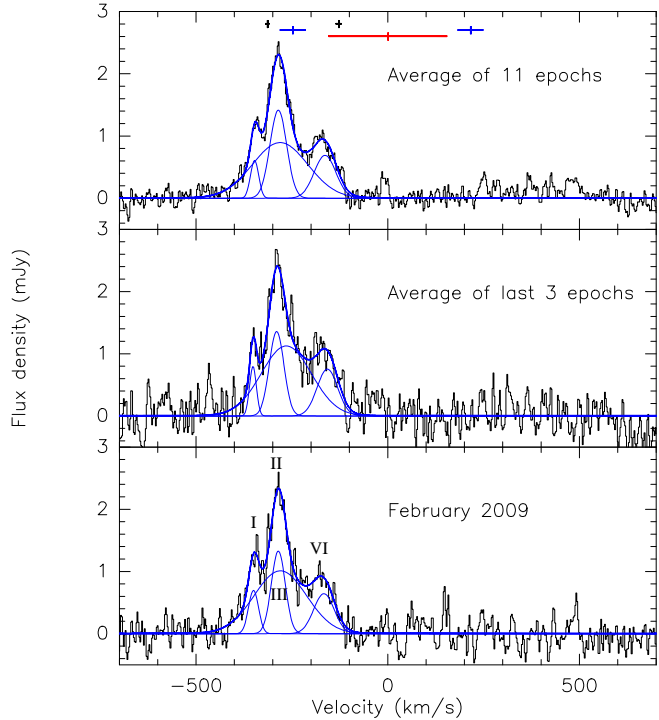} 
 \caption{{\it Lower panel}: Water maser spectrum of MG~J0414+0534 observed in February 2009. The labels {\sc I} to {\sc IV} indicate the best fit Gaussian components. {\it Middle panel}: Average of the last three epochs (September and November 2009 and January 2010) obtained using equal weights. {\it Upper panel}: Final spectrum produced by averaging all the epochs with the same weight. Individual Gaussian profiles fitted to the spectra are overlaid in blue together with the resulting profile. The channel spacing is 2.4\,km\,s$^{-1}$. The r.m.s noise level is 0.2\,mJy per channel in the spectra of the lower and middle panels and 0.1\,mJy per channel in the upper panel.}
   \label{fig:fit4}
\end{center}
\end{figure}

A significant change in the line profile seems to have occurred between the Effelsberg and EVLA observations and the first epoch of the Arecibo monitoring campaign. Indeed, the line appears to be much broader in the first Arecibo spectrum (taken in October 2008) w.r.t. the previous observations. The full width at half maximum (FWHM) of the Gaussian profile fitted to the line is 174$\pm$5\,km\,s$^{-1}$, i.\,e. more than a factor of two larger than the FWHM of the Effelsberg and EVLA spectra (\cite[Impellizzeri et al. 2008]{impellizzeri08}). As a consequence, we measure an unlensed isotropic luminosity of 26,000\,L$_{\odot}$, that makes the maser in MG~J0414+0534 the most luminous that is currently known. Furthermore, in October 2008 we tentatively detected a weaker satellite line at +470\,km\,s$^{-1}$ (Fig.\ref{fig:fig_oct}) that, however, was not confirmed by the spectra of the other epochs. This second feature, detected with a signal-to-noise ratio of three, is displaced by about 800\,km\,s$^{-1}$ from the main line and is five time less luminous. In February 2009 we performed deeper observations aimed at confirming the presence of this feature. No emission line other than the main one at the velocity of about $-$300\,km\,s$^{-1}$ was detected above a 3$\sigma$ noise level of 0.3\,mJy per 19.2\,km\,s$^{-1}$ channel. However, a weak feature is seen in the spectrum  at the velocity of about +490\,km\,s$^{-1}$ (see Fig.\ref{fig:fit4}, lower panel). The satellite line remains undetected also in the spectrum produced by averaging all of the epochs with the same weights (Fig~\ref{fig:fit4}, upper panel). Nonetheless, we note that the range between 200 and 500\,km\,s$^{-1}$ looks spiky and that, interestingly, one of these spikes is at the position of the satellite line. Averaging the spectra using different weights (e.\,g. 1/r.m.s$^2$ or the integration time) does not change the shape of the resulting spectrum significantly. This may indicate that many weak lines are present in the range 200--500\,km\,s$^{-1}$ and that in October 2008 we saw one of these lines flaring.  

The high SNR of the February 2009 spectrum ($\sim$13; see Fig~\ref{fig:fit4}, lower panel) reveals that the main line has a complex profile that is likely the result of the blending of at least four components with line widths between 30 and 160\,km\,s$^{-1}$. In order to inspect the variability of the individual velocity features, we produced a spectrum by averaging with equal weights the last three epochs of the monitoring campaign (September and November 2009 and January 2010). The resulting spectrum (Fig~\ref{fig:fit4}, middle panel) has an r.m.s comparable with that of the February 2009 observation. Comparing the Gaussian peak velocities, we find that the velocities of components I and II did not change, while the velocities of components III and IV have marginally increased by +15$\pm$3\,km\,s$^{-1}$ and +10$\pm$3\,km\,s$^{-1}$, respectively. Since velocity drifts have been observed only in very narrow lines (FWHM $\sim$ 1--4\,km\,s$^{-1}$), we think that the change in the peak velocities of these features maybe due to variations of a large number of sub-components, simulating a change in the radial velocity as observed most notably in NGC~1052 (\cite[Braatz et al. 1996]{braatz96}).

\section{Discussion and concluding remarks} 
Our monitoring data are partially consistent with our initial hypothesis that the emission is associated with the prominent relativistic jets of the quasar (\cite[Impellizzeri et al. 2008]{impellizzeri08}). First of all, even when the maser line profile is resolved into multiple velocity components, individual emission features have line widths between 30 and 160\,km\,s$^{-1}$ that resemble those of known H$_2$O masers associated with radio jets (e.\,g. Mrk~348; \cite[Peck et al. 2003]{peck03}). Our non-detection of a radial acceleration of the main maser peak is also compatible with the jet-maser scenario. 
The extreme stability of the main line peak and the continuum flux density in MG~J0414+0534 resulting from our study,  seems to exclude a jet-maser scenario similar to that in Mrk~348 (\cite[Peck et al. 2003]{peck03}) and NGC~1068 (\cite[Gallimore et al. 2001]{galli01}), while the reported significant variations in the line profile of our target may hint at similarities with the case of NGC~1052 (\cite[Braatz et al. 1996 and 2003]{braatz96,braatz03}). We note, however, that the number of sources in which the maser emission is confidently associated with the jet(s) is very low and that more of these masers should be studied in detail in order to investigate the properties of these kind of sources.
Furthermore, the tentative detection of the redshifted feature in the October 2008 spectrum is compatible with the disk-maser hypothesis. 
If the main maser line and the satellite line at +470\,km\,s$^{-1}$ are considered as the blueshifted and redshifted lines from the tangentially seen part of an edge-on accretion disk in Keplerian rotation, then the radius at which the emission originates is given by $R= GM_{\rm BH}V_{\rm R}^{-2}$, where $G$ is the gravitational constant, $M_{\rm BH}$ is the black hole mass, and $V_{\rm R}$ is the rotational velocity at radius $R$. From the difference between the line of sight velocities of the main and satellite maser lines ($V_{\rm obs}$), we obtain $V_{\rm R} = V_{\rm obs} \cdot \sin (i)^{-1}$ $\sim 370 \cdot \sin (i)^{-1}$\,km\,s$^{-1}$. Adopting the black hole mass of $M_{\rm BH}=10^{9.0}$\,M$_{\odot}$ calculated by \cite[Pooley et al. (2007)]{pooley07} for MG~J0414+0534, and assuming an edge-on orientation ($i$ = 90$^{\circ}$) for the accretion disk, we get a radius of $R \sim$ 30\,pc. This value is fairly large compared to the radii at which maser emission is found in the accretion disks of nearby radio quiet AGN (typically, 0.1 to 1 pc). We should keep in mind however, that MG~J0414+0534 is a radio loud quasar, while known disk-maser hosts are mainly radio quiet Seyfert or LINER galaxies with a mass of the nuclear engine that is two orders of magnitude lower ($\sim 10^7$\,M$_{\odot}$; \cite[Kuo et al. 2011]{kuo2011}). 

In conclusion, although we have been able to provide useful elements to determine the nature of the maser in MG~J0414+0534, our current data are presently insufficient to confidently rule out whether the maser emission is due to a jet-cloud interaction or a rotating circumnuclear disk. VLBI observations and longer time-scale single-dish monitoring will be essential to shed light on the origin of the H$_2$O maser in this distant quasar.

\end{document}